\documentclass[preprint,12pt,authoryear]{elsarticle}
\usepackage{lineno}
\usepackage{bm}
\usepackage[ruled,linesnumbered]{algorithm2e}
\usepackage{booktabs} 
\usepackage{siunitx} 
\usepackage[table]{xcolor}
\usepackage[utf8]{inputenc}
\usepackage{amssymb}
\usepackage{color}
\usepackage[left=2.5cm,right=2.5cm,top=3cm,bottom=3cm]{geometry}
\usepackage{svg}
\usepackage{amsmath}

\usepackage{multirow}

\usepackage{booktabs}
\usepackage{graphicx}
\usepackage{subfigure}
\usepackage{epstopdf}
\usepackage{picinpar}
\usepackage{amsmath}
\usepackage{amsfonts}
\usepackage[mathscr]{euscript}
\usepackage{calligra}
\usepackage{array}
\usepackage[authoryear]{natbib} 
\usepackage{setspace}
\usepackage{setspace}
\usepackage{lipsum} 
\usepackage{adjustbox}

\makeatletter
\def\ps@pprintTitle{} 
\makeatother

\usepackage{titlesec}
\titlespacing{\section}{0pt}{6pt}{6pt}
\titlespacing{\subsection}{0pt}{3pt}{3pt}
\titlespacing{\subsubsection}{0pt}{2pt}{2pt}

\DeclareMathAlphabet{\mathpzc}{OT1}{pzc}{m}{it}

\newdefinition{assumption}{Assumption}

\newdefinition{definition}{Definition}
\newdefinition{remark}{Remark}
\newproof{proof}{Proof}

\biboptions{authoryear,compress}

\let\OLDthebibliography\thebibliography
\renewcommand\thebibliography[1]{
  \OLDthebibliography{#1}
  \setlength{\parskip}{0pt}
  \setlength{\itemsep}{0pt plus 0.3ex}
}
\usepackage[colorlinks=true,
            linkcolor=blue,
            citecolor=blue,
            urlcolor=blue]{hyperref}  

\usepackage{cleveref}

\begin{document}

\begin{frontmatter}

\title{RESPOND: Risk-Enhanced Structured Pattern for LLM-driven Online Node-level Decision-making}

\makeatletter
\def\@author{\normalsize\@author}
\def\@address{\small\@address}
\makeatother

\author[a]{Dan Chen}
\ead{dan.chen@139.com}

\author[b]{Heye Huang\corref{cor1}}
\ead{heyeh@mit.edu}

\author[d]{Tiantian Chen}
\ead{nicole.chen@kaist.ac.kr}

\author[d]{Zheng Li}
\ead{zli2674@wisc.edu}

\author[e]{Yongji Li}
\ead{dragonlyj@gmail.com}

\author[f]{Yuhui Xu}
\ead{xuyuhui@gd.chinamobile.com}

\author[d]{Sikai Chen}
\ead{sikai.chen@wisc.edu}

\address[a]{School of Vehicle and Mobility, Tsinghua University}
\address[b]{Singapore-MIT Alliance for Research and Technology (SMART), Massachusetts Institute of Technology}
\address[c]{Cho Chun Shik Graduate School of Mobility, Korea Advanced Institute of Science and Technology}
\address[d]{Department of Civil and Environmental Engineering, University of Wisconsin-Madison}
\address[e]{Sun Yat-sen University}
\address[f]{China Mobile Group Guangdong Co., Ltd.}

\cortext[cor1]{Corresponding author}

%

\begin{abstract}
Current LLM-based driving agents relying on unstructured plain-text memory suffer from low-precision scene retrieval and inefficient reflection. To address this, we present RESPOND, a structured decision-making framework for LLM agents grounded in risk patterns. RESPOND constructs a unified $5\times3$ ego-centric matrix that encodes spatial topology and road constraints, enabling consistent retrieval of spatial--risk configurations. Building on this, a hybrid Rule+LLM pipeline employs a two-tier memory lookup: exact patterns ensure rapid, safe action reuse in high-risk contexts, while sub-patterns facilitate personalized style adaptation under low risk. Furthermore, a pattern-aware reflection mechanism abstracts tactical corrections from crash frames to update structured memory, achieving "one-crash-to-generalize" learning.
Experimental results demonstrate the effectiveness of this design. In \texttt{highway-env}, RESPOND surpasses state-of-the-art LLM-based and RL-based agents while generating substantially fewer collisions. With step-wise human feedback, the agent acquires a \emph{Sporty} driving style within approximately 20 decision steps through sub-pattern abstraction. For real-world validation, we evaluate RESPOND on 53 high-risk cut-in scenarios extracted from the HighD dataset. For each event, we intervene at the moment immediately preceding the cut-in and allow RESPOND to re-decide the driving action. Compared to the recorded human behavior, RESPOND’s decisions reduce subsequent risk in 84.9\% of these scenarios, demonstrating the feasibility and practical relevance of RESPOND under real-world driving conditions. These results highlight the potential for real-world autonomous driving, personalized driving assistance, and proactive hazard mitigation. Code will be released at: \url{https://github.com/gisgrid/RESPOND}.
\end{abstract}

\begin{keyword}
Large Language Models (LLM); Risk Pattern Representation; Structured Memory; Reflection Learning; Autonomous Driving
\end{keyword}

\end{frontmatter}

%
\section{Introduction}
\label{sec:introduction}

Achieving safe and reliable autonomous driving in complex mixed-traffic environments requires decision-making systems that can retrieve past experiences precisely, adapt to new situations, and learn efficiently from sparse safety-critical events\citep{huang2025lead, wang2021towards, zhao2025saca}. Recent work has explored Large Language Models (LLMs) for driving decision-making, leveraging their reasoning and generalization capabilities to interpret diverse traffic situations \citep{mao2023gpt, zhou2026safedrive, wen2023road}. Despite these advances, current LLM-driven agents primarily store experiences as plain-text narratives, which limits their utility in several fundamental ways. First, scene retrieval becomes imprecise because semantic matching alone cannot reliably distinguish small but safety-critical variations in high-risk or long-tail events. Second, reflection learning is inefficient since crash cases stored in unstructured text make it difficult to extract reusable knowledge, restricting generalization from failures. Moreover, most existing approaches lack explicit encoding of spatial--risk dependencies between the ego vehicle and surrounding entities, making it difficult to represent the risk-perception patterns that human drivers rely on when making safety-critical decisions.

To address these limitations, we propose RESPOND, a knowledge-driven decision framework grounded in the quantitative Driver Risk Field (DRF) \citep{kolekar2020human,huang2020probabilistic}. RESPOND encodes each ego-centric scene as a unified $5\times3$ risk matrix that captures neighboring vehicles and road-edge constraints, yielding a discrete and comparable spatial--risk pattern for reliable memory retrieval and improved recall fidelity over plain-text storage. Based on this representation, RESPOND introduces a hybrid Rule+LLM decision pipeline with a two-tier memory structure: exact pattern matches enable high-confidence action reuse in high-risk contexts, while sub-pattern matches summarize essential spatial--risk configurations to support personalization under low risk \citep{kou2025padriver,zeng2025adrd}. A lightweight LLM refines decisions only when necessary, preserving interpretability and adaptability. In addition, a pattern-aware reflection mechanism enables one-crash-to-generalize learning by extracting corrected strategies from pre- and post-crash frames and writing them back as reusable sub-patterns, preventing repeated failures in unseen scenarios.
The main contributions are summarized as follows:
\begin{enumerate}
    \item We propose a unified risk--environment representation that encodes ego-centric traffic context into a DRF-based 5×3 risk matrix. This structured formulation yields a discrete and comparable pattern space that enables high-precision memory retrieval and improves over unstructured textual storage.
    \item We design a two-tier decision framework that combines rule-based strategies with LLM reasoning. Exact pattern reuse supports high-confidence actions in safety-critical scenarios, while sub-pattern abstraction enables flexible decision-making and efficient human-in-the-loop personalization under low risk.
    \item We develop a pattern-aware reflection mechanism that analyzes pre- and post-crash frames to extract tactical corrections and writes them back into structured memory. This mechanism supports one-crash-to-generalize learning and accelerates knowledge accumulation when safety-critical data are limited.
\end{enumerate}

The remainder of this paper is organized as follows. Section~2 reviews related works. Section~3 outlines the RESPOND framework. Section~4 details the methodology. Section~5 presents experiments and results on both simulation and real-world datasets. Section~6 concludes the paper and discusses limitations and future research directions.

%
\section{Related Works}
\label{rw}
%
\subsection{Quantitative Risk Modeling}
Risk modeling serves as a critical foundation for ensuring safe and context-aware autonomous decision-making. Existing approaches can be broadly categorized into three classes.

(1) \textbf{Kinematics-based surrogate safety indicators.}
Surrogate safety metrics such as Time-To-Collision (TTC), Post-Encroachment Time (PET), and Deceleration Rate to Avoid Collision (DRAC) provide interpretable and computationally efficient approximations of short-term collision risk using vehicle trajectory data \citep{munro2008threshold, peesapati2018can, cheng2025emergency, shalev2017formal}. These indicators are widely used in traffic conflict analysis and trajectory-level risk assessment. However, they are inherently limited to pairwise interactions and often fail to capture multi-agent spatial structures, relative topologies, or cooperative risk contexts, especially in dense or highly interactive traffic.

(2) \textbf{Probabilistic graphical risk models.}
Probabilistic models such as Bayesian networks and Partially Observable Markov Decision Processes (POMDPs) offer a principled framework for risk estimation under uncertainty \citep{lauri2022partially, spaan2012partially, ross2011bayesian}. These methods can model hidden states, incorporate noisy observations, and evaluate long-horizon, policy-aware risk. Despite their theoretical advantages, they often suffer from computational complexity, the curse of dimensionality, and limited scalability in real-time multi-agent environments, especially when planning over dense spatiotemporal grids.

(3) \textbf{Potential field and gradient-based risk methods.} Classical Artificial Potential Field (APF) and Social Force Models \citep{bounini2017modified, helbing1995social} define attractive and repulsive forces for path planning or pedestrian–vehicle interaction. While intuitive and lightweight, they are prone to local minima, lack semantic integration, and are typically unsuitable for behavior reasoning. More recent advancements formulate risk as structured scalar or vector fields that integrate semantic, physical, and kinematic cues. Representative among these is the DRF model \citep{kolekar2020human}, which encodes ego-relative risk intensity across multiple directions by combining distance, relative velocity, geometry, and regulatory constraints. Variants such as the Driving Safety Field \citep{wang2016driving, huang2020probabilistic} extend these ideas to multi-source risk fusion.
DRF-based fields are well suited for structured encoding because they naturally produce spatial grid representations of interaction risk that can be converted into comparable matrix patterns for decision memory. However, most existing risk models are stateless, assessing risk only instantaneously without capturing temporal evolution, action–response dynamics, or feedback-driven correction. Moreover, they lack mechanisms for storing and retrieving pattern-level risk–response associations, which limits their applicability in memory-augmented decision frameworks.

\subsection{LLM-Based Driving Agents}
LLMs have recently gained significant traction in autonomous driving due to their ability to interpret complex situations, reason over action sequences, and produce human-readable justifications \citep{yang2023llm4drive, wang2025learning}. Existing work in this domain can be grouped into three primary directions.

(1) \textbf{Natural language-based scene understanding and high-level decision making.}
Early studies explore LLMs for processing scene descriptions and generating high-level actions such as lane changing or yielding \citep{mao2023gpt, wen2023road}. Multimodal extensions such as DriveLM and Driveagent \citep{wang2023drivemlm, cui2024survey, hou2025driveagent} incorporate BEV or vision inputs to condition reasoning. While these systems demonstrate strong explainability and zero-shot generalization, most rely on free-text prompts and responses, making it difficult to align or reuse decisions across structurally similar scenes. As a result, they suffer from semantic drift, inconsistent action formulations, and limited support for robust decision recall in safety-critical situations.

(2) \textbf{Memory-augmented and reflection-capable LLM agents.}
Recent progress integrates memory and self-reflection into LLM-driven control. DiLu \citep{wen2023dilu} employs closed-loop analysis of past failures to refine subsequent outputs. General-purpose reflection systems such as Reflexion and Self-Refine \citep{zhou2026safedrive, cai2024driving} show that LLM agents can improve long-horizon task performance through iterative feedback without retraining. Despite their effectiveness, these systems store memory in unstructured natural language, causing retrieval to rely mainly on semantic similarity rather than structural correspondence. This limitation prevents detection of textually dissimilar yet structurally similar high-risk cases, hindering fine-grained reflection and pattern-level recall.

(3) \textbf{Personalized autonomous driving and human-in-the-loop adaptation.}
Personalization in AV decision-making has gained increasing interest. Systems such as PADriver and Drive as you say \citep{kou2025padriver, cui2024receive} use user instructions, preference descriptions, or demonstrations to modulate driving style, enabling cautious or aggressive behavior tuning. These approaches leverage LLM flexibility for style modulation across risk conditions. However, they typically operate at the trajectory or policy level without a structured abstraction of personalized behavior. Consequently, style transfer remains scenario-dependent and cannot be encoded into reusable decision structures.

%
\begin{figure}[htbp]
\centering
\includegraphics[width=1\textwidth]{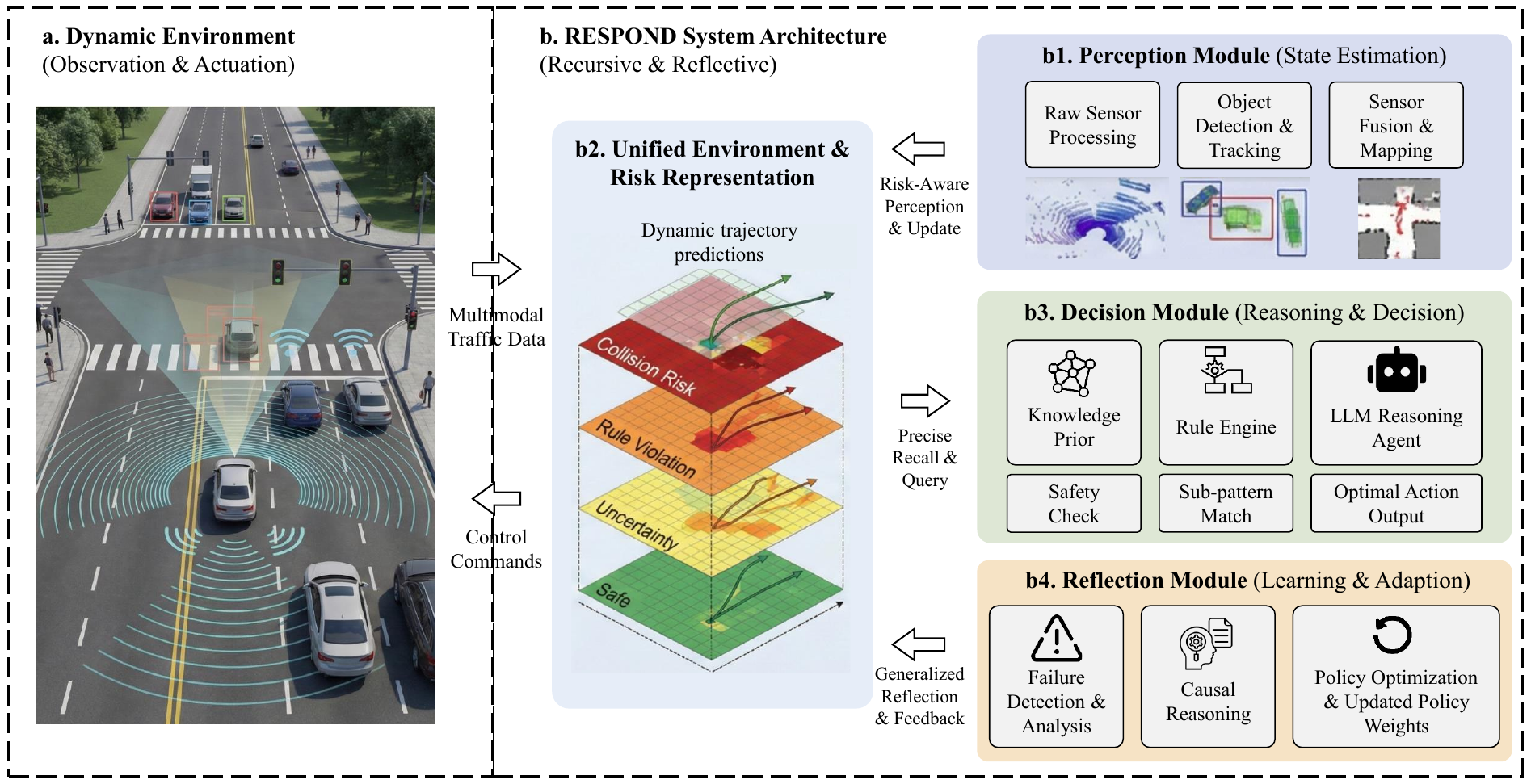}
\caption{Architecture of the RESPOND framework. 
(a) Interaction between the autonomous agent and the dynamic environment through perception, decision, and action. 
(b) Recursive and reflective system architecture with four core modules: (b1) Perception Module for state estimation; (b2) Unified environment and risk representation for encoding multi-layer spatial and interaction risk semantics; (b3) Decision module that integrates knowledge priors, deterministic rule engines, and LLM-based reasoning to generate safe actions; and (b4) Reflection module for failure analysis and policy optimization. The closed-loop design enables continual safety improvement through experience accumulation and reflection.}
\label{fig:fig1}
\end{figure}

\section{Framework}
\label{sec:framework}

The RESPOND framework is designed to overcome the limitations of unstructured textual memory in LLM-based agents, which often suffer from low retrieval fidelity and inefficient reflection. By introducing a unified risk-aware architecture, RESPOND integrates structured perception, hybrid reasoning, and closed-loop learning to enable precise context recognition and rapid adaptation. As illustrated in Fig.~\ref{fig:fig1}, the architecture transforms the ego-centric scene into a structured format through the Perception Module (b1) and the Unified Environment \& Risk Representation (b2). The perception module processes raw sensor data through object detection and fusion to estimate states. These states are then mapped into a multi-layered representation encompassing collision risk, rule violation, and uncertainty. This process culminates in a unified $5\times3$ spatial--risk matrix, which discretizes the ego-centric surroundings to enable consistent pattern retrieval and bridge the gap between physical dynamics and symbolic reasoning.

Building on this structured representation, the Decision Module (b3) adopts a hybrid pipeline to balance safety and flexibility. A knowledge prior and deterministic rule engine enable sub-pattern matching to reuse high-confidence actions in familiar scenarios, ensuring stable and interpretable behavior. For novel or complex contexts where predefined rules are insufficient, an LLM-based reasoning agent is activated to generate candidate actions, which are subsequently validated through a safety checking mechanism. The Reflection Module (b4) closes the learning loop by detecting safety-critical events, conducting causal analysis to identify failure factors, and updating policy weights in the structured memory. This process enables one-crash-to-generalize learning, allowing the system to adapt from limited failures while reducing the recurrence of similar unsafe behaviors.


\section{Methodology}
\label{sec:method}

This section first formalizes a unified risk–environment encoding that maps heterogeneous sensor inputs into a structured cognitive state. It then presents the hierarchical memory for efficient retrieval, followed by the hybrid decision pipeline and pattern-aware reflection that support continuous safety improvement and personalization.

\subsection{Unified Risk--Environment Encoding}
\label{subsec:encoding}

To bridge the gap between physical risk modeling and language-driven reasoning, RESPOND introduces a Unified Risk--Environment Encoding. This strategy transforms heterogeneous driving data, spatial topology, dynamic states, and quantitative risk, into a compact, discrete representation optimized for structured memory storage and LLM inference.

\subsubsection{Hierarchical Abstraction for Structured Scene Representation}
\label{sec:abstraction}

\begin{figure}[t]
    \centering
    \includegraphics[width=1\textwidth]{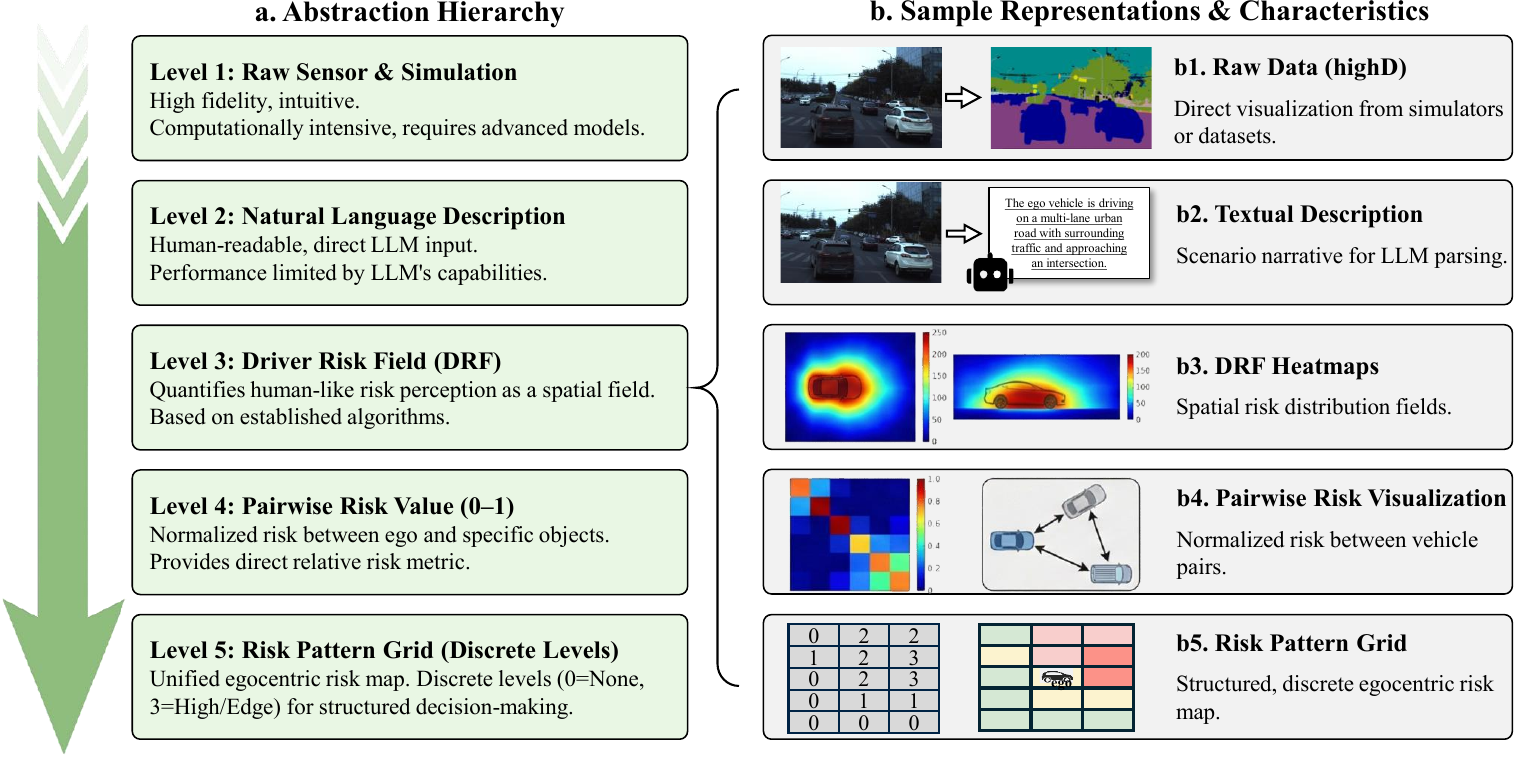}
    \caption{Five-level abstraction hierarchy for structured scene and risk representation. (a) Progressive abstraction from raw sensory inputs to discrete risk patterns. (b) Representative outputs at each level: (b1) raw sensor or simulation data, (b2) natural language description, (b3) continuous DRF heatmaps, (b4) pairwise normalized risk values, and (b5) the final unified $5 \times 3$ discrete risk pattern grid.}
    \label{fig:fig2}
\end{figure}

Existing LLM-based driving agents face a fundamental mismatch between semantic flexibility and quantitative rigor. While text-based descriptions are expressive and LLM-friendly, they lack the numerical precision required for safety-critical reasoning. In contrast, classical DRF models provide physically grounded risk quantification but do not offer a structured and retrievable format suitable for memory-augmented LLM inference. To bridge this gap, we design a five-level abstraction hierarchy (Fig.~\ref{fig:fig2}) that progressively distills high-dimensional driving data into a compact and structured Risk Pattern.
The hierarchy starts from Level~1 (Raw Sensor \& Simulation), which captures high-fidelity vehicle states from simulators or real-world datasets such as HighD. Level~2 (Natural Language Description) translates these states into narrative text that is directly consumable by LLMs, albeit with limited efficiency and consistency. Level~3 (Driver Risk Field) maps kinematic states into continuous DRF heatmaps, providing a spatial visualization of interaction risk. Level~4 (Pairwise Risk Value) further summarizes these fields into normalized scalar risk values in the range $[0,1]$, quantifying directional interactions between the ego vehicle and surrounding objects. Finally, Level~5 yields the Risk Pattern Grid, a unified $5 \times 3$ discrete matrix that encodes spatial topology and quantified risk intensity in a fixed-format representation. This final abstraction serves as the core unit for structured memory storage and subsequent decision-making.

\subsubsection{Quantitative Risk Modeling}
We adopt the DRF model as the quantitative basis for Level~4 and extend its formulation to omnidirectional ego-centric interactions beyond forward-facing obstacles. For a surrounding vehicle $j$, the risk value $RV_j$ is defined as the normalized overlap between the risk fields of the ego vehicle and vehicle $j$:

\begin{equation}
    RV_{j} = \frac{A_{\text{overlap}}(\text{ego}, j)}{A_{\text{ref}}}
\end{equation}
where $A_{\text{overlap}}$ denotes the intersection area of the two risk fields projected onto the road plane. The normalization term $A_{\text{ref}}$  is selected according to the relative interaction direction: 
\begin{equation}
A_{\text{ref}} =
\begin{cases}
A_{\text{ego}}, & \text{for front and front-adjacent interactions} \\
A_{j}, & \text{for rear and rear-adjacent interactions}
\end{cases}
\end{equation}
Here, $A_{\text{ego}}$ denotes the area of the ego vehicle’s risk field, and $A_{j}$ denotes that of the surrounding vehicle. This direction-aware normalization captures asymmetric interaction dynamics, particularly in rear-following scenarios where risk is dominated by the trailing vehicle. To account for multi-lane interactions, the ego DRF is laterally shifted to compute risk values in six directional zones: front, rear, left-front, left-rear, right-front, and right-rear.

\subsubsection{Discrete Matrix Representation}
To transform continuous risk values into a retrieval-ready structure, we discretize the spatial context into a grid matrix $P \in \{0, 1, 2, 3\}^{5 \times 3}$. The matrix is centered on the ego vehicle, covering the ego lane and two adjacent lanes (left and right), with five longitudinal segments extending from rear to front. Each element $P_{i,j}$ in the matrix represents the risk level of a specific spatial cell, discretized based on the computed $RV$:

\begin{equation}
    P_{i,j} = \begin{cases} 
    0, & \text{if } RV_{i,j} < 0.34 \quad (\text{Safe}) \\
    1, & \text{if } 0.34 \le RV_{i,j} < 0.66 \quad (\text{Attention}) \\
    2, & \text{if } 0.66 \le RV_{i,j} < 0.99 \quad (\text{Danger}) \\
    3, & \text{if } RV_{i,j} = 1 \quad (\text{Critical})
    \end{cases}
\end{equation}

To capture hard safety constraints that may be underestimated by pure DRF in low-velocity scenarios, we introduce two deterministic override rules. First, road-edge constraints are enforced by assigning $P_{i,j}=1$ to cells corresponding to non-drivable regions, explicitly encoding road topology into the risk pattern. Second, near-field risk in the center column of the $5\times3$ matrix is encoded using proximity and TTC rather than DRF overlap. If no surrounding vehicle lies within a predefined distance threshold (30~m in our experiments), the cell value is set to 0; otherwise, TTC is mapped to discrete risk levels $\{1,2,3\}$.
The resulting $5\times3$ risk pattern integrates road layout and interaction risk into a compact, standardized representation. Compared with raw sensor data, this formulation substantially reduces state dimensionality while preserving critical safety information, enabling precise memory retrieval and stable LLM reasoning.

\subsection{Structured Memory Construction}
\label{subsec:memory}

\begin{figure}[t]
\centering
\includegraphics[width=1\textwidth]{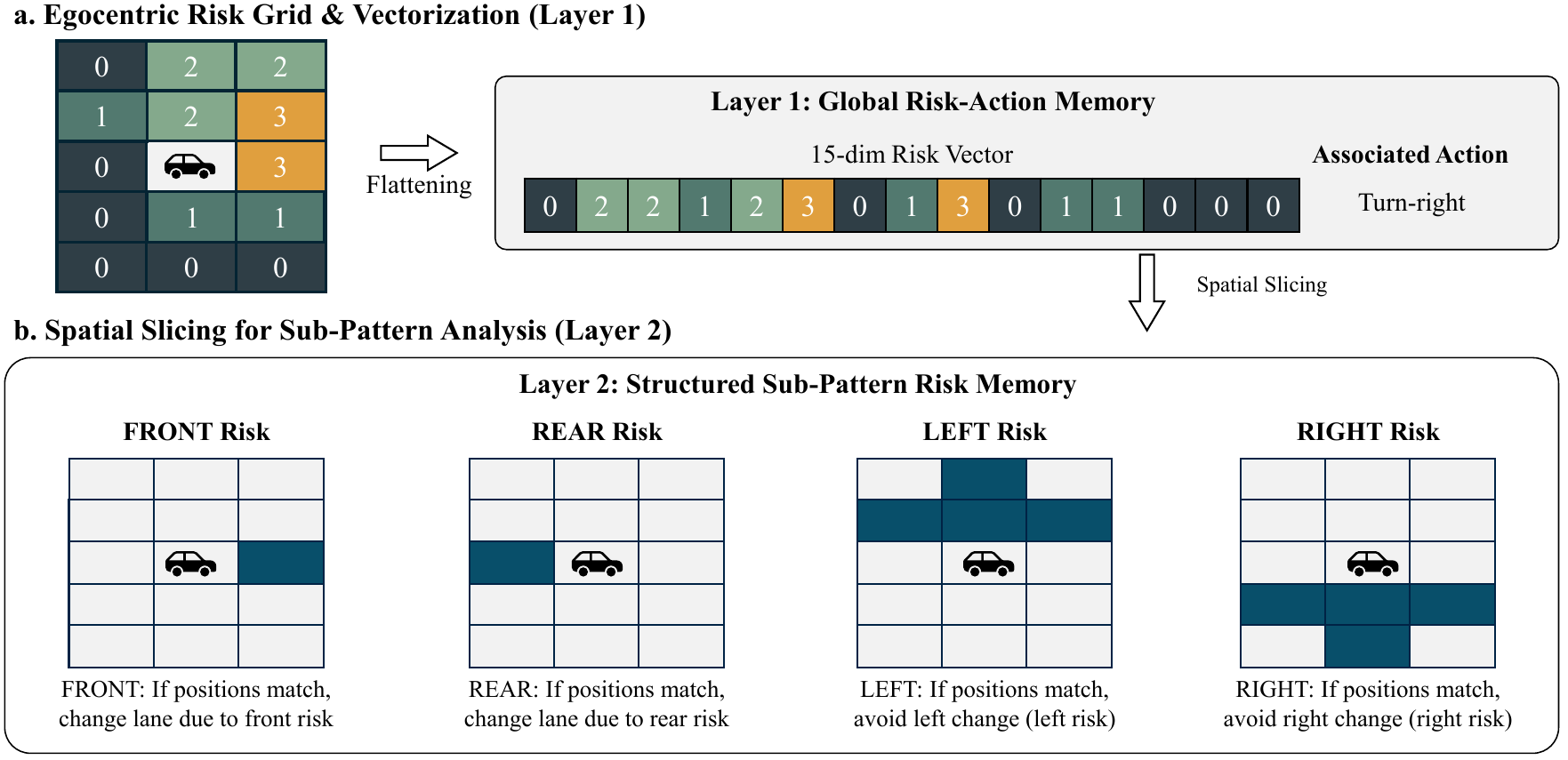}
\caption{Structured memory architecture. (a) Layer 1: Global Risk-Action Memory. The risk grid is flattened into a 15-dimensional vector for exact global retrieval. (b) Layer 2: Structured Sub-Pattern Risk Memory via Spatial Slicing. The grid is sliced into four tactical sub-patterns: FRONT/REAR triggers active strategies (e.g., change lane), while LEFT/RIGHT imposes safety constraints (e.g., avoid lane change).}
\label{fig:fig3}
\end{figure}

To support risk-aware decision reuse, RESPOND builds a two-layer structured memory system (Fig.~\ref{fig:fig3}). Layer 1, the \textit{Global Risk-Action Memory}, flattens the $5 \times 3$ risk grid into a 15-dimensional vector paired with an executed action. This vectorization enables high-precision exact matching for familiar global scenes. To address data sparsity and enable generalization, Layer 2, the \textit{Structured Sub-Pattern Risk Memory}, employs \textit{spatial slicing} to decompose the grid into semantic sub-patterns. Specifically, FRONT and REAR patterns capture longitudinal risks to guide strategic lane changes, while LEFT and RIGHT patterns identify lateral risks that serve as negative constraints (e.g., avoid unsafe maneuvers). This dual structure enables the agent to replicate validated actions in high-risk contexts while flexibly applying tactical rules in novel scenarios.

\subsubsection{Layered Memory Architecture}

At the core of RESPOND’s memory module is a flattened form of the risk pattern matrix in Section~\ref{subsec:encoding}. Each $5 \times 3$ risk map is reshaped into a 15-dimensional vector $\bm{v} \in \{0,1,2,3\}^{15}$ and paired with the corresponding action label $\mathcal{A}$.

\textbf{Layer 1: Exact Pattern Memory.} The first layer stores a collection of such tuples $\mathcal{M}_{1} = \{(\bm{v}_{k}, \mathcal{A}_{k})\}_{k=1}^{N}$. This layer enables fast nearest-neighbor retrieval via Euclidean distance minimization:
\begin{equation}
    (\bm{v}^*, \mathcal{A}^*) = \arg \min_{(\bm{v}_{k}, \mathcal{A}_{k}) \in \mathcal{M}_{1}} \| \bm{v}_{\text{obs}} - \bm{v}_{k} \|
\end{equation}
where $\bm{v}_{\text{obs}}$ is the current observation vector. This vectorized structure eliminates ambiguity introduced by linguistic variation in text-based memories and supports deterministic recall of previously validated decisions. To further improve data efficiency, RESPOND exploits inherent left--right lane symmetry: whenever a risk vector is inserted, its horizontally mirrored counterpart is automatically generated and added, effectively doubling knowledge density without additional data collection.

\textbf{Layer 2: Sub-Pattern Abstraction Memory.} Although full-pattern matching is precise, the combinatorial space of risk vectors ($4^{15}$) leads to sparsity in unseen environments. Layer~2 addresses this issue by storing sub-patterns, i.e., abstracted spatial fragments that capture critical local configurations. We define four canonical types: (i) FRONT / REAR, which focus on central longitudinal cells and indicate proximity threats requiring strategic lane changes; and (ii) LEFT / RIGHT, which focus on lateral columns and act as constraints (e.g., an occupied lane) that suppress corresponding maneuvers. These sub-patterns provide human-interpretable abstractions, enabling tactical knowledge transfer across structurally similar but globally distinct scenes.

\subsubsection{Hierarchical Retrieval and Decision Support}
During deployment, the memory system operates through a prioritized retrieval mechanism to balance precision and generality:

\textbf{Exact Reuse (Layer 1):} The system first queries Layer 1. If an exact match ($\|\bm{v}_{\text{obs}} - \bm{v}^*\| = 0$) is found with high confidence, the corresponding action $\mathcal{A}^*$ is reused directly. This ensures consistency and trustworthiness for previously validated high-risk scenarios.

\textbf{Strategic Inference (Layer~2):} When no exact match is found, Layer~2 sub-patterns are queried to provide high-level intent (e.g., a FRONT match suggesting a lane change) or to constrain the action space (e.g., a LEFT match blocking leftward maneuvers). These strategic cues are passed to the LLM to refine the final action.

The two-layer structured memory follows three core design principles. 
(i) Compression and directionality: high-dimensional scene data are mapped to low-dimensional risk vectors, filtering irrelevant noise while preserving task-relevant spatial topology; sub-patterns further focus on decision-critical regions. 
(ii) Symmetry and augmentation: geometric mirroring exploits physical invariants in driving to increase data efficiency and coverage. 
(iii) Abstraction for generalization: Layer~1 enables exact recall, while Layer~2 extends coverage through symbolic abstraction, supporting reflection-based updates at both levels. This design allows RESPOND to internalize corrections after a single failure, achieving one-crash-to-generalize learning and providing a structured foundation for decision-making and continual adaptation.

\subsection{Hybrid Decision-Making Pipeline}
\label{subsec:decision}

While structured memory enables precise retrieval, autonomous driving also requires a decision-making engine that balances the safety guarantees of rule-based systems with the adaptability of LLMs. To address this trade-off, RESPOND employs a hybrid decision-making pipeline that combines explicit risk-aware constraints with contextual LLM reasoning, dynamically switching strategies based on a quantified risk metric.

\subsubsection{Risk Quantification and Regime Classification}
The pipeline first quantifies the urgency of the current scene. We define the instantaneous risk level, $RL_t$, as the maximum risk intensity in the longitudinal zones, derived from the unified risk encoding:
\begin{equation}
    RL_t = \max(RV_{\text{front}}, RV_{\text{rear}})
\end{equation}
We establish a safety threshold $\tau = 0.75$. The decision logic bifurcates into two regimes: the Safety-Critical Regime when $RL_t \ge \tau$, prioritizing collision avoidance; and the Efficiency-Oriented Regime when $RL_t < \tau$, focusing on driving style and comfort.

\subsubsection{Safety-Critical Regime (High Risk)}
\begin{figure}[t]
\centering
\includegraphics[width=1\textwidth]{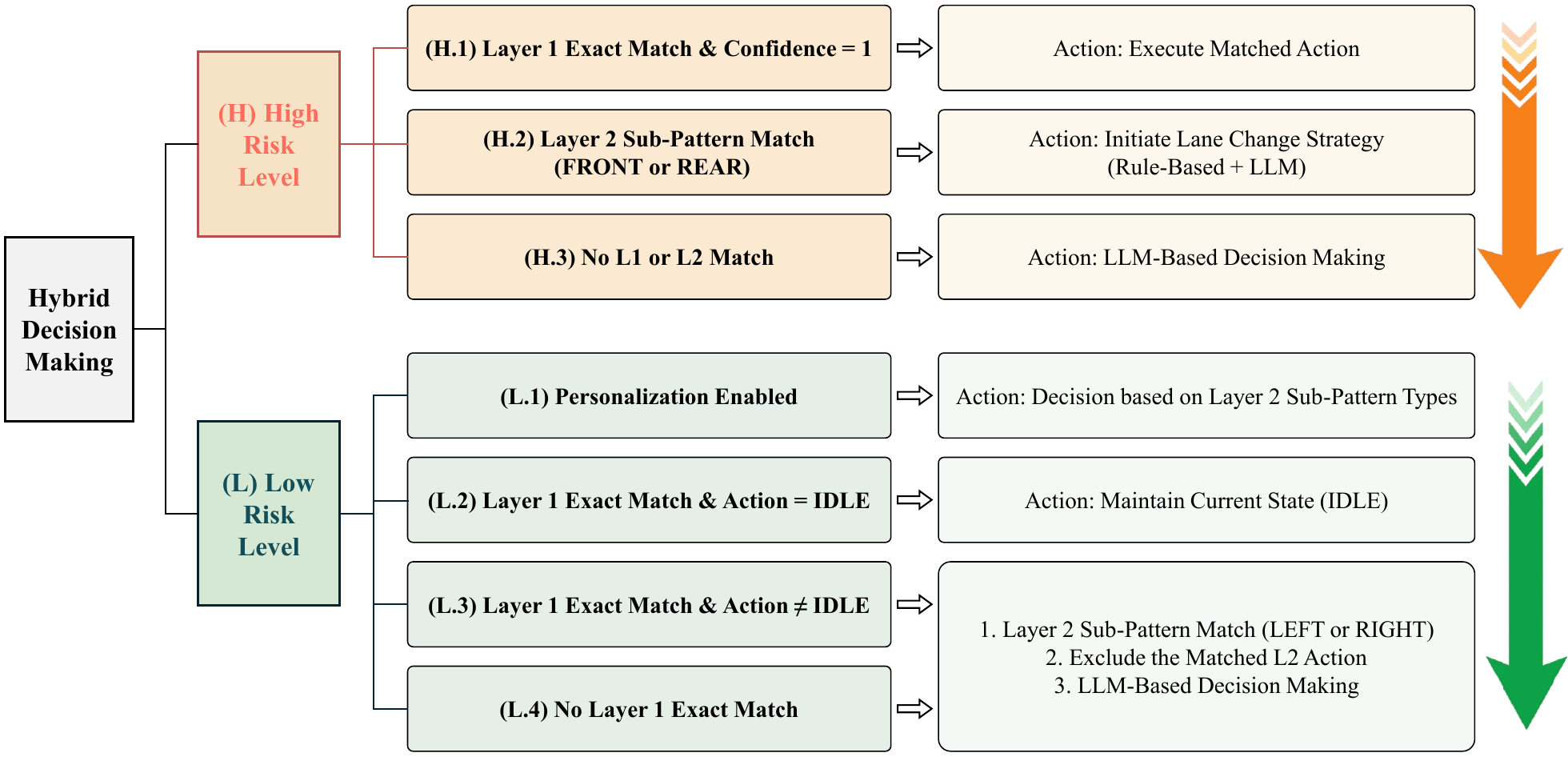}
\caption{Hybrid decision-making logic in the RESPOND framework. The system first classifies the scene as high- or low-risk based on DRF-derived risk values. For high-risk scenes, it prioritizes exact pattern reuse (H.1), sub-pattern strategies with LLM refinement (H.2), or zero-shot LLM fallback (H.3). For low-risk contexts, RESPOND enables personalized style guidance (L.1), IDLE action reuse (L.2), filtered action selection with lateral constraints (L.3), or default LLM reasoning (L.4). This modular architecture ensures interpretable, efficient, and risk-adaptive driving behavior.}
\label{fig:fig4}
\end{figure}

As illustrated in the upper branch of Fig.~\ref{fig:fig4}, in high-risk contexts ($RL_t \ge 0.75$), the system follows a three-tier logic to ensure deterministic safety.
(H.1) Exact Pattern Reuse: The system first queries Layer~1 memory $\mathcal{M}_1$. If an exact match with verified safety confidence (Confidence = 1) is found, the associated action $a^*$ is executed directly ($a_t = a^*$ if $\|\bm{v}_{\mathrm{obs}} - \bm{v}^*\| = 0$). This step bypasses LLM reasoning, ensuring a validated response with minimal latency.
(H.2) Sub-Pattern Constrained Reasoning: If no exact match is available, the system retrieves Layer~2 sub-patterns, such as FRONT or REAR interactions. A matched sub-pattern activates a strategic intent $S$ (e.g., change lane). Deterministic rules $\mathcal{R}$ then restrict the candidate action space $\mathcal{A}$ to a safe subset $\mathcal{A}_{\mathrm{safe}}$, from which a lightweight LLM selects the final action.
(H.3) Zero-Shot Fallback: If no memory match is found, the system resorts to zero-shot reasoning. The LLM is prompted with structured risk values $\mathbf{RV}$ to generate a conservative action, where quantitative risk priors are used to guide reasoning and reduce unsafe or implausible trajectories.

\subsubsection{Efficiency-Oriented Regime (Low Risk)}
Conversely, in low-risk environments ($RL_t < 0.75$), the system optimizes for personalization and computational efficiency, following the lower branch of Figure~\ref{fig:fig4}:
(L.1) Personalized Adaptation: If personalization is enabled, the system retrieves stylistic sub-patterns from Layer 2 (e.g., $P_{\text{style}}$) to guide action selection, aligning behavior with user preferences (e.g., Sporty vs. Comfort).
(L.2) Idle Optimization: If a Layer 1 match suggests an \texttt{IDLE} action, it is executed directly. Since steady-state driving constitutes the majority of operation time, this shortcut significantly reduces average inference cost.
(L.3/L.4) Constraint-Aware Reasoning: For non-idle actions or novel scenarios, Layer 2 lateral sub-patterns (\textit{LEFT/RIGHT}) serve as negative constraints. Specifically, if a lateral risk is detected ($RV_{\text{lat}} > \tau_{\text{lat}}$), corresponding lane-change actions are masked from the LLM's output space, ensuring that even generative reasoning remains within safety boundaries.

\subsection{Pattern-Aware Reflection Mechanism}
\label{subsec:reflection}

Although the hybrid decision pipeline enables adaptive reasoning, traditional learning methods (e.g., RL or end-to-end imitation) typically require thousands of episodes to converge on stable behaviors. In contrast, human drivers rapidly internalize failures, often a single accident suffices to prevent recurrence. RESPOND replicates this efficiency through a Pattern-Aware Reflection Mechanism (RMR). By coupling quantitative risk patterns with symbolic sub-patterns, RMR facilitates rapid credit assignment and memory updating, achieving a ``one-crash-to-generalize'' capability.

\begin{figure}[h]
    \centering
    \includegraphics[width=1\textwidth]{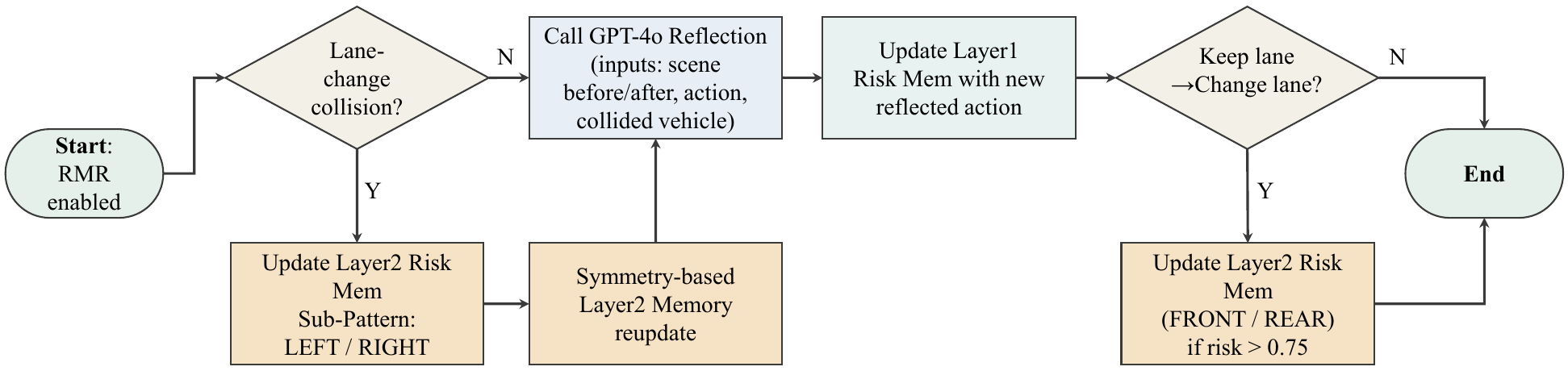}
    \caption{The Pattern-Aware Reflection Mechanism (RMR). Upon detecting a failure or near-miss event, the system activates a reflection loop. For lane-change collisions, lateral constraints are directly inferred and used to update Layer-2 sub-patterns via geometric symmetry. For more complex collisions, a high-capacity LLM analyzes pre- and post-crash context to identify causal factors and synthesize corrective actions. The resulting insights are abstracted into reusable sub-patterns and written back to structured memory, enabling avoidance of similar failures in future scenarios.}
    \label{fig:fig5}
\end{figure}

\subsubsection{Structured Reflection Workflow}
The reflection process is triggered upon detecting a collision or safety violation (see Fig.~\ref{fig:fig5}). The mechanism distinguishes between two failure modes to apply targeted corrections:

\textbf{Lane-Change Collisions (Lateral Updates):} If a collision results from a lateral maneuver, the system immediately infers a spatial constraint. A new \textit{LEFT} or \textit{RIGHT} sub-pattern is recorded in Layer 2, explicitly marking that direction as unsafe under the current risk configuration. Leveraging geometric symmetry, RESPOND automatically mirrors this constraint to the opposite side to double learning efficiency.

\textbf{General Collisions (LLM-Based Analysis):} For complex scenarios, the system invokes a high-capacity LLM (e.g., GPT-4o) with a structured prompt containing pre- and post-collision frames, the executed action, and the collision object's state. The LLM performs causal reasoning to identify the root cause and synthesizes a revised action $a^*$ (e.g., ``decelerate'' instead of ``maintain speed'').

\subsubsection{Memory Evolution and Strategic Abstraction}
To close the learning loop, the insights derived from reflection are written back into the structured memory. The revised action $a^*$ is first stored in Layer 1 memory with a confidence score of 1. Crucially, if the failure reveals a higher-level tactical shift, e.g., switching from \textit{keep lane} to \textit{change lane}, and the pre-collision risk satisfies $RV_{\text{front}} \text{ or } RV_{\text{rear}} \ge 0.75$, a corresponding strategic sub-pattern (FRONT or REAR) is updated in Layer 2. This abstraction transforms specific numeric failures into generalized safety heuristics.

The reflection mechanism follows three key principles: 
(i) Hierarchical credit assignment, where Layer~1 stores precise vector–action corrections and Layer~2 captures abstract strategies to support both fine-grained and generalized reuse; 
(ii) symmetry-driven augmentation, which exploits spatial symmetry to automatically mirror lateral sub-patterns (LEFT/RIGHT) without additional data, yielding a $2\times$ gain in learning efficiency; and 
(iii) strategy-level abstraction, which focuses on tactical semantics rather than specific collision geometries, enabling transfer to structurally similar but unseen risk configurations.

In summary, RMR introduces a structured post-collision learning loop that bridges LLM-based causal inference with symbolic memory rewriting. By integrating dual-layer updates for both exact and abstract representations, the system achieves ``one-crash-to-generalize'' learning. This mechanism enables RESPOND to continually evolve its risk perception and decision strategies, effectively closing the perception--decision--reflection--memory loop and ensuring robust adaptation in high-risk, long-tail scenarios.

\subsection{Human-in-the-Loop Personalization}
\label{subsec:personalization}

While the reflection mechanism enables autonomous improvement from failures, capturing subjective driving preferences (e.g., Sporty vs. Comfort) requires explicit human guidance. RESPOND addresses this need with a human-in-the-loop personalization module built on Layer~2 memory, allowing driving behavior to be shaped through lightweight interactive feedback rather than large-scale retraining, thereby enabling data-efficient personalization without compromising safety.

\begin{figure}[h]
    \centering
    \includegraphics[width=1\textwidth]{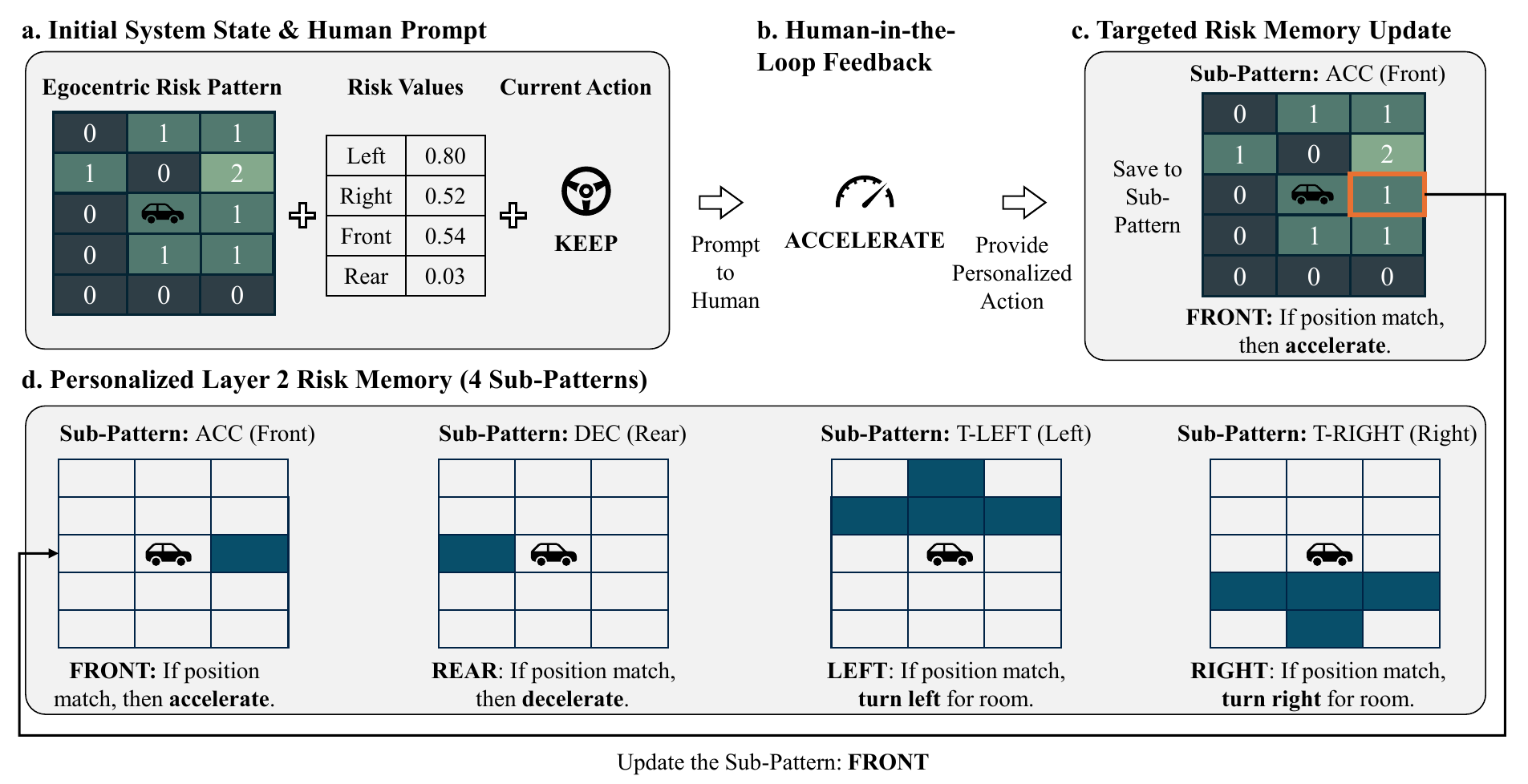}
    \caption{The Human-in-the-Loop Personalization workflow. (a) The system presents the current ego-centric risk pattern and quantified risk values alongside its proposed action. (b) The human driver provides feedback (e.g., ``Accelerate'') based on their preference. (c) RESPOND abstracts this feedback into a personalized Layer 2 sub-pattern (e.g., associating a specific front risk level with acceleration) and updates the memory, enabling the reproduction of this driving style in future low-risk scenarios.}
    \label{fig:personalization}
\end{figure}

\subsubsection{Interactive Learning Workflow}
The personalization process follows a three-step cycle (Fig.~\ref{fig:personalization}). First, in personalization mode, the system presents a structured summary of the current context, including the $5\times3$ risk pattern and surrounding risk values. Second, the user evaluates the proposed action and provides a preferred alternative (e.g., Accelerate instead of Keep Lane). Finally, RESPOND abstracts this feedback into a stylistic sub-pattern (e.g., if front risk $<0.6$, then Accelerate) and stores it in Layer~2 under the corresponding profile. During deployment, when the risk level is low ($RL<0.75$), these personalized sub-patterns are prioritized to guide action selection in structurally similar contexts.

\subsubsection{Efficiency and Safety Constraints}
This design adheres to three core principles:
(i) Safety-Constrained Adaptation, where user preferences are strictly gated by the quantitative risk threshold ($RL < 0.75$), ensuring that personalization never overrides safety-critical responses;
(ii) Abstract-Level Learning, which stores preferences as generalized rules rather than raw sensory data, allowing the system to internalize a consistent style from as few as $\sim$20 feedback instances; and
(iii) Scalability, as the symbolic nature of Layer 2 sub-patterns enables the seamless transfer of personalized profiles across different vehicles or platforms. By uniting human intuition with structured knowledge representation, RESPOND achieves rapid, verifiable, and safe customization of autonomous driving behavior.

RESPOND’s human-in-the-loop module enables interpretable and risk-aware personalization by embedding user preferences directly into the sub-pattern space, allowing fine-grained behavioral adjustment without retraining. Through sparse feedback, the system efficiently captures driver-specific styles while enforcing safety by activating personalization only under $RL < 0.75$. The structured sub-pattern memory also preserves persistent user traits, offering portable and reusable behavior templates for scalable deployment in consumer AVs and robotaxi platforms. This design integrates symbolic reasoning, structured pattern encoding, and human feedback to achieve a safety-aligned and user-tailored driving experience.

\section{Experiments}
\label{sec:experiments}

This section validates the RESPOND through closed-loop simulation, personalization experiments, and real-world evaluation. The experiments examine three hypotheses: (i) structured risk–environment encoding and memory improve safety and generalization; (ii) reflection-driven learning enables one-crash-to-generalize efficiency; and (iii) the system robustly transfers to real-world human driving data in a zero-shot setting \citep{huang2020integrated}.

\subsection{Experimental Setup and Metrics}
\label{subsec:setup}

Two complementary platforms were adopted: the \textit{highway-env} simulator \citep{leurent2018environment} for controlled evaluation under configurable traffic densities, and the \textit{HighD} dataset \citep{krajewski2018highd} for validating safety reasoning against real-world human trajectories.

\textbf{Implementation Details.}
All simulations were implemented in Python 3.9 using PyTorch 2.5 for the numerical computation of the risk field. The OpenAI GPT-4o-mini API was employed for real-time decision reasoning, while the high-capacity GPT-4o was reserved for post-collision reflection. In \textit{highway-env}, we followed the protocol of DiLu, conducting episodes of 30 decision steps (1 Hz) with varying lane configurations (4--5 lanes) and traffic densities (2.0--3.0).

\textbf{Evaluation Metrics.}
To quantitatively assess performance, we utilized the following metrics: (i) Success Rate (SR): the percentage of episodes completed without collision; (ii) Learning Efficiency: the performance gain per reflection iteration; (iii) Risk Reduction ($\Delta R$): the average reduction in cumulative DRF intensity; and (iv) Human Evaluation Rate: the ratio of human experts judging RESPOND’s decisions as safer than recorded human actions in real-world scenarios.

\subsection{Comparative Analysis in Closed-Loop Simulation}
\label{subsec:simulation_results}

To evaluate the effectiveness of RESPOND in controlled driving scenarios, we conducted extensive closed-loop experiments in the \textit{Highway-env} simulator. This platform enables configurable multi-lane highways with adjustable traffic densities, serving as a rigorous benchmark for tactical decision-making. Following the protocol of DiLu, each evaluation episode lasts 30 seconds (1 Hz decision frequency), involving dynamic interactions with surrounding agents sampled from empirical distributions.

We benchmark RESPOND against three representative baselines: DiLu (LLM-based reasoning with textual memory), GRAD (SOTA RL-based method trained on 600,000 episodes), and PADriver (personalized MLLM-based agent). Evaluations were performed across three increasing difficulty levels: \textit{Lane-4 (Density 2.0)}, \textit{Lane-5 (Density 2.5)}, and \textit{Lane-5 (Density 3.0)}. The quantitative results are summarized in Table~\ref{tab:success_rate}.

\begin{table}[h]
\centering
\caption{Driving Success Rate (SR \%) across different traffic densities.}
\label{tab:success_rate}
\renewcommand{\arraystretch}{1.2}

{\small
\begin{tabular}{lcccc}
\hline
\textbf{Method} & \textbf{Lane-4 (2.0)} & \textbf{Lane-5 (2.5)} & \textbf{Lane-5 (3.0)} & \textbf{Decline (pp)} \\ \hline
GRAD & 69\% & 47\% & 10\% & $\downarrow$ 59 \\
DiLu & 70\% & 65\% & 35\% & $\downarrow$ 35 \\ 
RESPOND L1 (Ours) & 85\% & 75\% & 50\% & $\downarrow$ 35 \\ 
\textbf{RESPOND L1+L2 (Ours)} & \textbf{100\%} & \textbf{80\%} & \textbf{70\%} & \textbf{$\downarrow$ 30} \\ \hline
RESPOND Sporty & 80\% & - & - & - \\ \hline
\end{tabular}
}

\vspace{0.35em}
\begin{minipage}{0.92\linewidth}
\footnotesize \textbf{Notes:}
Decline denotes the performance drop from the standard setting (Lane-4, 2.0) to the most complex setting (Lane-5, 3.0).
\end{minipage}

\end{table}

\textbf{Performance Superiority and Robustness.}
As shown in Table~\ref{tab:success_rate}, RESPOND (L1+L2) achieves a 100\% Success Rate in the standard setting, significantly outperforming both DiLu (70\%) and GRAD (69\%). Even the simplified version using only Layer 1 memory (RESPOND L1) reaches 85\% SR, demonstrating the inherent advantage of structured risk encoding over unstructured text or pure RL policies. More critically, RESPOND exhibits superior robustness in complex, unseen environments. When transferring from 4-lane training to high-density 5-lane evaluation (Density 3.0), the performance of GRAD collapses to 10\% (a 59 percentage point drop), and DiLu declines to 35\%. In contrast, RESPOND (L1+L2) maintains a robust 70\% SR, with the smallest relative decline (30 pp). This resilience confirms that the Layer 2 sub-pattern abstraction captures transferable strategic knowledge (e.g., generic lane-change logic) that remains valid even when global scene configurations change drastically.

\textbf{Learning Efficiency.}
Learning efficiency is a critical bottleneck for deploying autonomous agents. The contrast in sample efficiency is striking:
(i) GRAD requires 600,000 RL episodes to converge to 69\% SR;
(ii) DiLu requires approximately 40 reflection iterations to reach 70\% SR;
(iii) RESPOND, by leveraging symmetry augmentation and one-crash-to-generalize abstraction, requires only 6 collision events to learn 16 reusable memory units and achieve 100\% SR.
This represents an orders-of-magnitude gain in sample efficiency, validating that structured memory enables rapid acquisition of safety-critical patterns.

\textbf{Personalization Capability.}
Additionally, the \textit{RESPOND Sporty} mode, trained with only 20 steps of human feedback (detailed in Section~\ref{subsec:personalization_exp}), achieves an 80\% SR in the standard setting with an average speed of 19.39 m/s. This performance is competitive with PADriver Normal (20.13 m/s) but requires significantly less training overhead, demonstrating that RESPOND can effectively balance safety with personalized driving styles.

\subsection{Efficiency of Reflection-Based Learning}
\label{subsec:reflection_efficiency}

A critical bottleneck in autonomous driving is the high cost of acquiring safety knowledge from sparse failure events. Unlike reinforcement learning (RL) which requires millions of interactions, or textual reflection methods like DiLu which need repetitive failures to stabilize, RESPOND aims to achieve ``one-crash-to-generalize'' learning. We systematically evaluated this efficiency by conducting controlled experiments in the \textit{highway-env} simulator, starting with empty memory banks.

\begin{table}[h]
\centering
\caption{Reflection efficiency and generalization performance.}
\label{tab:reflection_efficiency}
\renewcommand{\arraystretch}{1.2}

{\small
\begin{tabular*}{\linewidth}{@{\extracolsep{\fill}} lccc}
\hline
\textbf{Metric} & \textbf{DiLu} & \textbf{RESPOND L1} & \textbf{RESPOND L1+L2} \\ \hline
\textbf{Collisions Required} & $\sim40$ & \textbf{12} & \textbf{6} \\
\textbf{Memories per Crash} & 1 (Text) & 1.7 (Vector) & \textbf{2.6 (Vector + Sub)} \\
\textbf{Collision Efficiency Gain} & $1\times$ & {$\sim3\times$} & \textbf{$\sim7\times$} \\ \hline
Generalization Rate: & & & \\
FRONT/REAR Types & $\approx30\%$ & 57\% & \textbf{100\%} \\
LEFT/RIGHT Types & $\approx30\%$ & 57\% & \textbf{97\%} \\ \hline
\end{tabular*}
}

\vspace{0.4em}
\begin{minipage}{0.92\linewidth}
\footnotesize \textbf{Notes:}
Collisions Required is the number of failures needed to reach a 70\% success rate, and Generalization Rate is the percentage of recurring accidents eliminated after one reflective update.
\end{minipage}

\end{table}

\textbf{Learning Throughput and Efficiency.}
Table~\ref{tab:reflection_efficiency} quantifies the learning throughput. To accumulate sufficient experience for stable driving (reaching 70\% SR), DiLu required approximately 40 collisions. In contrast, RESPOND (L1+L2) achieved a 100\% SR after only 6 collisions, corresponding to a {$\sim7\times$} improvement in collision-level learning efficiency. This efficiency stems from two structural advantages:
(i) Symmetry Augmentation: Leveraging the spatial symmetry of the $5\times3$ risk matrix, RESPOND automatically generates mirrored memory entries for lateral conflicts (e.g., left lane change risks imply right lane constraints), doubling data utility.
(ii) Sub-Pattern Abstraction: A single failure allows the extraction of both an exact Layer 1 pattern and multiple Layer 2 sub-patterns (e.g., a generic ``Front Blocked'' rule). As a result, 6 collisions produced 16 reusable memory units, enabling rapid policy stabilization with significantly fewer failure events than DiLu.

\textbf{One-Shot Generalization Capability.}
We further assessed the ``one-crash-to-generalize'' hypothesis by forcing specific collision types (e.g., aggressive cut-ins) and measuring recurrence in structurally similar but numerically distinct scenarios (varying lane counts or traffic densities). As shown in Table~\ref{tab:reflection_efficiency}, RESPOND (L1+L2) eliminated 100\% of repeated longitudinal (FRONT/REAR) collisions and 97\% of lateral (LEFT/RIGHT) collisions after a single reflective update. 
In comparison, the exact-match-only variant (RESPOND L1) eliminated only 57\% of recurrences and failed under spatial variation. DiLu performed worse ($\approx$30\%), as textual memories poorly generalized across geometric differences (semantic similarity $<0.1$). These results confirm that Layer~2 sub-patterns act as generalized safety priors, enabling inference of avoidance strategies across families of risk configurations from a single example.

\textbf{Impact on Overall Performance.}
The superior sample efficiency directly translates to closed-loop robustness. In the Lane-4 (2.0) setting, RESPOND reached a 100\% success rate after only 6 failures, whereas GRAD required 600{,}000 episodes to achieve 69\%, and DiLu plateaued at 70\% after 40 failures. Moreover, the memory write-back process is fully interpretable, with each update corresponding to a specific risk cell or symbolic rule (e.g., ``LEFT $\to$ Avoid''). Overall, RESPOND offers a data-efficient, interpretable, and safety-driven alternative to conventional RL and textual memory approaches.

\subsection{Personalization under Low-Risk Conditions}
\label{subsec:personalization_exp}

Beyond safety-oriented decision-making, RESPOND supports human-in-the-loop personalization under low-risk conditions through its Layer~2 sub-pattern memory. Unlike prior approaches such as PADriver, which require large-scale supervised training and GPU-intensive fine-tuning, RESPOND achieves efficient personalization using a 20-step human-guided episode without model modification. When personalization is enabled and the risk level remains low ($RL<0.75$), the system presents the current $5\times3$ risk pattern, directional DRF-based risk values, and the proposed action to the driver. If an alternative action is provided (e.g., ACCELERATE or TURN LEFT), RESPOND maps it to a predefined sub-pattern type and stores it in the corresponding user profile. These abstract preferences are reused in subsequent low-risk scenarios to guide decision-making.

To evaluate the effectiveness of this lightweight personalization mechanism, we compare RESPOND against PADriver’s pre-trained style variants (Slow, Normal, Fast) and DiLu under the lane-4-density-2.0 configuration. Table~\ref{tab:sr_speed_combined} reports the quantitative results across different driving styles. Despite operating at only 1 Hz, the baseline RESPOND L1+L2 achieves a perfect 100\% success rate, outperforming PADriver Slow (97\%) which benefits from a higher 10 Hz control frequency. The RESPOND Sporty mode, trained using only a single 20-step HITL session, achieves an 80\% success rate, nearly matching PADriver Normal (83\%) and significantly outperforming PADriver Fast (57\%). These results confirm that abstract Layer-2 preferences can encode meaningful user intent without critically compromising safety.

\begin{table}[htbp]
\centering
\small
\caption{Success Rate (SR) and Average Speed comparison under lane-4-density-2.0 setting.}
\label{tab:sr_speed_combined}
\renewcommand{\arraystretch}{1.2}
\begin{tabular}{lccc}
\hline
\textbf{Method} & \textbf{SR (\%)} & \textbf{Avg. Speed (m/s)} & \textbf{Notes} \\ \hline
DiLu & 93 & 12.85 & 10 Hz decision freq.; PADriver setting\\
PADriver Slow & 97 & 18.46 & Safety-focused finetuned agent \\
PADriver Normal & 83 & 20.13 & Balanced performance \\
PADriver Fast & 57 & 24.12 & High-speed, unsafe \\
\textbf{RESPOND L1+L2} & \textbf{100} & 16.90 & Hybrid rule-memory reasoning, 1 Hz \\
RESPOND Sporty & 80 & 19.39 & 20-step HITL personalization \\ \hline
\end{tabular}
\end{table}

Driving style preferences are further analyzed via average speeds. As shown in Table~\ref{tab:sr_speed_combined}, while PADriver Fast reaches the highest speed (24.12 m/s), it suffers from a prohibitive collision rate. RESPOND Sporty achieves an average speed of 19.39 m/s, closely approximating PADriver Normal (20.13 m/s), whereas the standard RESPOND L1+L2 maintains a conservative 16.90 m/s to maximize safety. DiLu exhibits the most conservative behavior at 12.85 m/s. These comparisons demonstrate that RESPOND is capable of replicating personalized driving behaviors that are both interpretable and efficient, using only a few sub-pattern-level rules to modulate the trade-off between aggressiveness and caution.

Regarding system cost and sample efficiency, RESPOND avoids data-heavy finetuning and operates with zero retraining, achieving orders-of-magnitude reduction compared with PADriver. Personalization is introduced only under low-risk conditions, ensuring that preference-driven adaptation does not compromise collision-critical decisions. Qualitative inspection of Sporty-mode trajectories shows intuitive behaviors, including selective acceleration, safe-space overtaking, and suppression of risky maneuvers under rising conflict levels. Overall, RESPOND delivers style-aware personalization with minimal supervision and computational load. It preserves safety, generalizes through abstract pattern encoding, and provides a low-cost mechanism for deploying user-tailored autonomous driving behaviors.

\subsection{Real-World Robustness Validation on HighD Dataset}
\label{subsec:highd_validation}

Simulation environments, while controllable, inherently lack the stochasticity and long-tail
edge cases characteristic of real human driving. To evaluate RESPOND under real-world
conditions, we integrate it with the highD dataset, a large-scale drone-captured collection
of naturalistic highway trajectories. This experiment assesses whether RESPOND’s hybrid
Rule+LLM architecture can robustly handle unseen, safety-critical scenarios in real driving
data, and whether the unified risk encoding allows effective decision-making through
structured risk abstraction, even in the absence of exact prior memory matches.

\begin{figure}[t]
\centering
\includegraphics[width=1\textwidth]{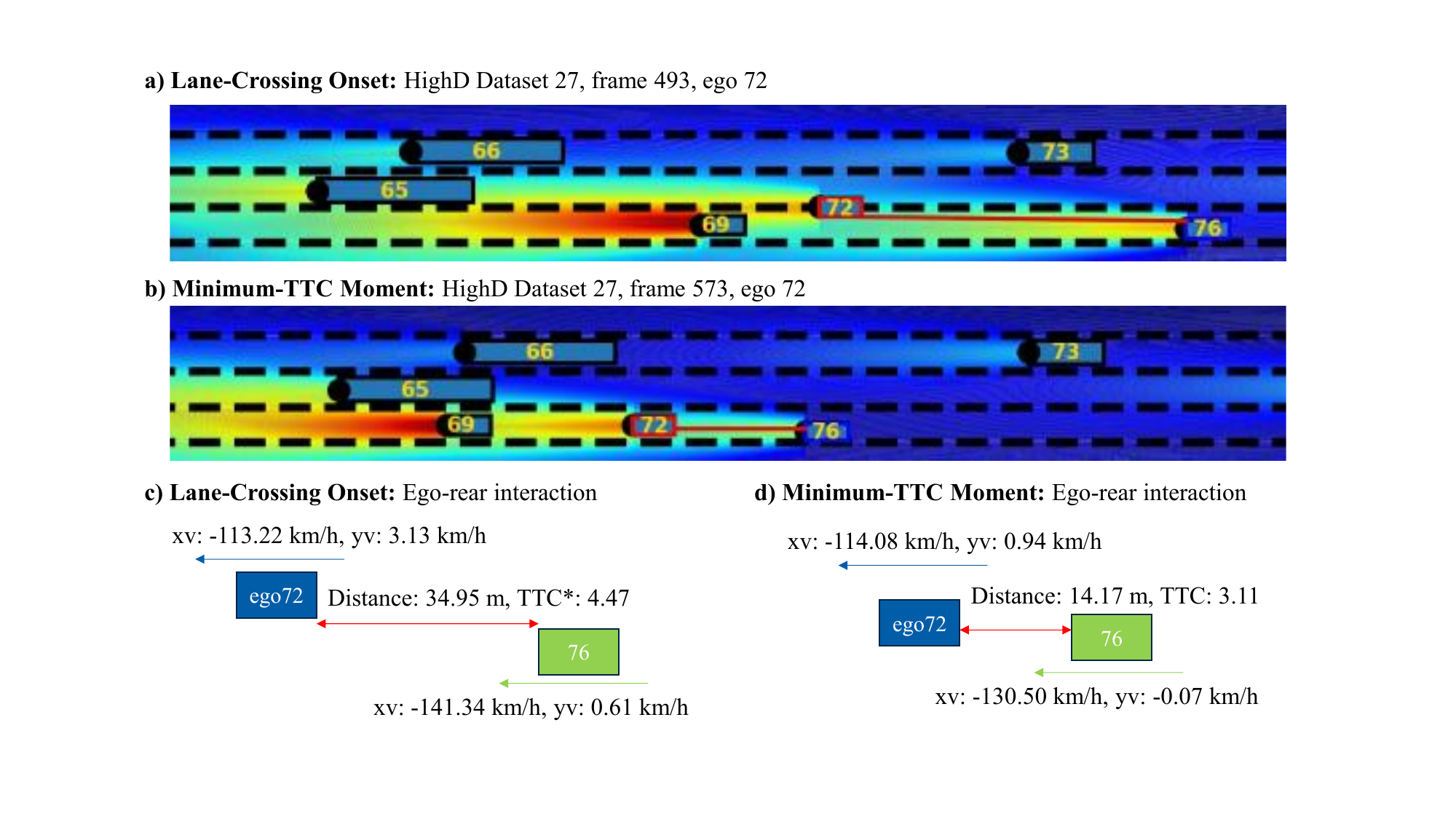}
\caption{Example of a safety-critical lane-change scenario from the HighD dataset.
(a–b) Global context views of ego vehicle 72 at two moments: (a) lane-crossing onset, when the ego vehicle starts crossing the lane boundary, and (b) the subsequent minimum-TTC moment, when the longitudinal TTC reaches its minimum after lane entry.
(c–d) Ego-centric schematics illustrating the interaction between the ego vehicle and the rear vehicle 76, including relative velocities, inter-vehicle distance, and the resulting TTC. Subfigures (c) and (d) correspond to the lane-crossing onset and minimum-TTC moment, respectively.
The TTC value in (c) is marked with an asterisk (*) because the longitudinal projections of the two vehicles have not yet overlapped along the lateral (y) axis at lane-crossing onset.}
\label{fig:fig7}
\end{figure}

\textbf{Dataset and Scenario Selection.} To assess real-world robustness, we systematically extract safety-critical lane-change scenarios from the complete highD dataset (Fig.~\ref{fig:fig7}  as an example). We first identify all lane-change events and treat the lane-changing vehicle as the ego agent. For each maneuver, we evaluate the longitudinal risk induced by the lane change with respect to the following vehicle in the target lane after the ego vehicle merges. A scenario is labeled as high-risk if the rear TTC falls below 4.0 s at any point following lane entry. Using this criterion, we extract a total of 53 high-risk real-world scenarios from the HighD dataset. For each selected event, we reconstruct the decision context at the moment immediately preceding the ego vehicle’s lane-change initiation, encode the scene into the unified $5 \times 3$ risk pattern, and compute four directional DRF-based risk values.

\textbf{Real-World Intervention Evaluation Protocol.}  This experiment evaluates whether RESPOND’s hybrid Rule+LLM architecture can produce safe and effective driving decisions when intervening in previously unseen, safety-critical scenarios derived from real-world driving data. Rather than relying on exact prior memory matches, RESPOND operates through structured risk abstraction and large language model reasoning to handle long-tail situations.

\begin{figure}[t]
\centering
\includegraphics[width=1\textwidth]{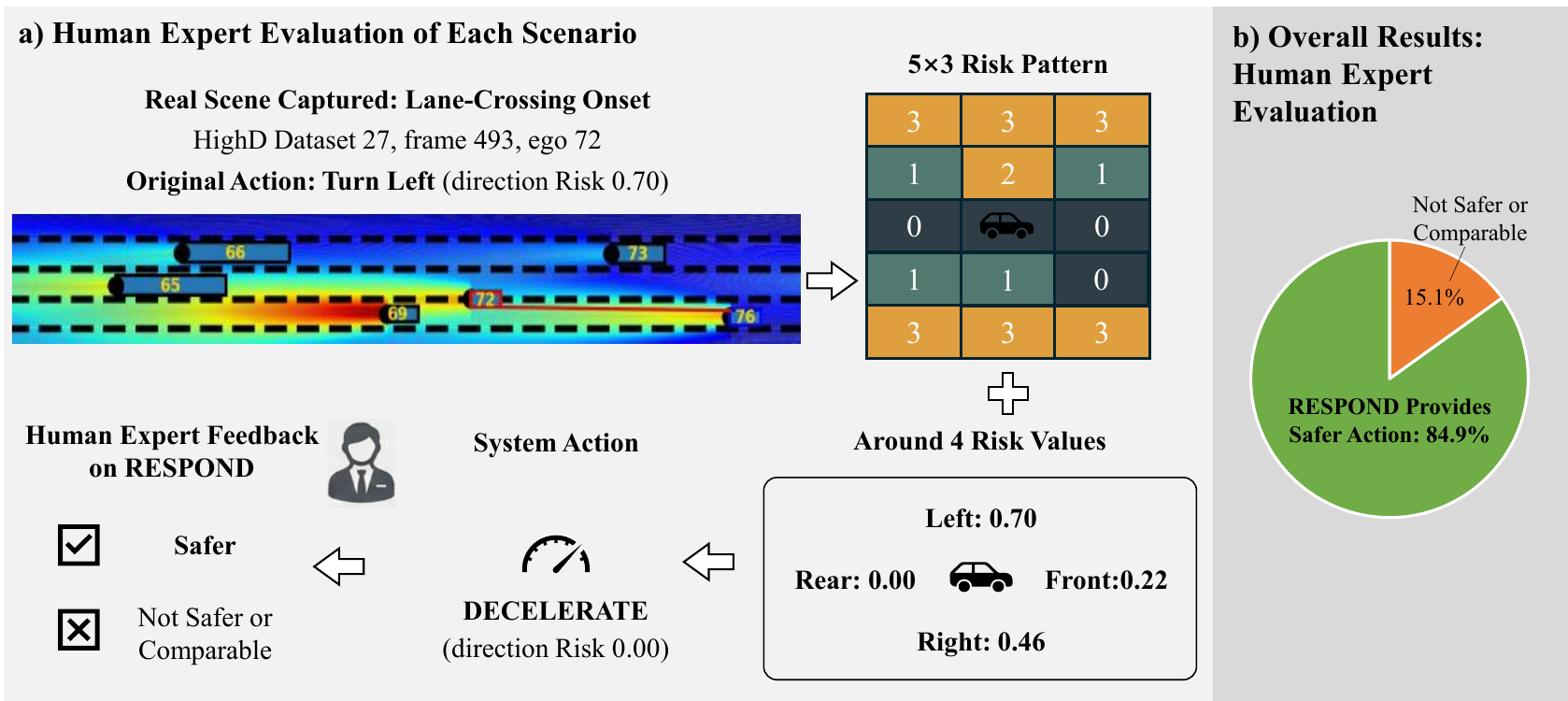}
\caption{Human expert evaluation of RESPOND on the HighD dataset.
(a) Evaluation protocol for safety-critical scenarios. Experts review the scene at lane-crossing onset, the original human action, RESPOND’s zero-shot action, the directional risk profile, and the video replay, and make judgments based on video context and domain experience.
(b) Results over 53 high-risk scenarios. RESPOND’s actions were judged safer in 84.9\% of cases; the remaining 15.1\% include 5 comparable and 3 worse than human actions.}
\label{fig:fig8}
\end{figure}

For each selected cut-in event, RESPOND is queried using a structured prompt encoding
the ego-centric spatial configuration of surrounding vehicles, quantified directional risk
values, and current motion states. Based on this input, the LLM generates a high-level
action suggestion (e.g., ``decelerate'' or ``change lane left''). To assess decision quality,
RESPOND’s actions are compared against the recorded human actions at the same
decision point. Safety is evaluated through expert human assessment informed by the
visual driving context and associated risk metrics  (Fig.~\ref{fig:fig8}).

\textbf{Quantitative Results and Analysis.}
The evaluation results are summarized in Table~\ref{tab:highd_results}. In the majority of cases, RESPOND's decisions were judged to be safer than the original human actions. Specifically, out of 53 high-risk scenarios, RESPOND provided a safer decision in 45 cases (84.9\%). Among the remaining cases, 5 scenarios (9.4\%) were assessed as comparable to human actions, while in 3 scenarios (5.7\%), the original human actions were judged to be safer.

It is worth noting that these results were obtained by directly applying the same RESPOND configuration used in the highway-env simulation without additional tuning for real-world data. Specifically, the prompting strategy was kept unchanged. In simulation, RESPOND is guided to consider the average speed of surrounding vehicles and maintain a comparable velocity, preventing overly conservative behavior that would otherwise result in unnecessarily slow driving to maximize safety and success rate. When transferred to real-world HighD scenarios, which predominantly consist of extreme cut-in events with short TTC, this design choice can occasionally lead RESPOND to favor acceleration in situations where surrounding vehicles exhibit unusually high speeds.

In addition, real-world traffic introduces vehicle characteristics not explicitly modeled in the simulator. In particular, the current risk field formulation primarily computes DRF based on the front point of each vehicle. For long vehicles such as trucks, common in HighD, this simplification does not fully account for the spatial extent of the vehicle body, potentially introducing discrepancies between the computed risk values and the actual perceived risk in certain scenarios.

These observations reflect known limitations of transferring a simulator-calibrated configuration to real-world data without adaptation. Our goal in using HighD is therefore not to demonstrate immediate deployability, but to assess whether RESPOND retains meaningful decision-making capability and practical relevance under real-world driving conditions.

\begin{table}[h]
\centering
\caption{Zero-shot evaluation results on high-risk HighD scenarios.}
\label{tab:highd_results}
\renewcommand{\arraystretch}{1.2}

{\small
\begin{tabular}{lcc}
\hline
\textbf{Comparison Outcome} & \textbf{Count} & \textbf{Percentage} \\ \hline
Human Action Safer& 3& 5.7\%\\
 Comparable& 5&9.4\%\\
\textbf{RESPOND Action Safer} & \textbf{45}& \textbf{84.9\%}\\ \hline
\textbf{Total Cases} & 53& 100\% \\ \hline
\end{tabular}
}

\vspace{0.4em}
\begin{minipage}{0.92\linewidth}
\footnotesize \textbf{Notes:}
Safer Decision Rate denotes the percentage of cases where expert reviewers judged RESPOND's action to be safer than the recorded human driver’s action in 53 high-risk HighD scenarios.
\end{minipage}

\end{table}
Qualitative analysis shows that RESPOND consistently leverages DRF-based risk gradients
to avoid high-risk driving behaviors, even without memory recall. In extreme high-speed and safety-critical scenarios, RESPOND suppresses unsafe lane changes under limited lateral
space and maintains lane discipline when relative speed differences with surrounding vehicles would lead to hazardous post–lane-change TTC conditions. These behaviors indicate that the unified risk-environment encoding grounded in DRF offers sufficient structural guidance for the LLM to reason about risk and generate well-controlled decisions in previously unseen environments.
The zero-shot results on HighD further highlight key architectural strengths. The seamless transfer from simulation to real data confirms the domain-invariant nature of the $5\times3$ risk representation, which requires no retraining. In addition, risk-quantified LLM reasoning effectively bridges physical dynamics and symbolic planning, enabling RESPOND to serve as a robust fallback policy for long-tail scenarios and a scalable tool for offline safety analysis and edge-case mining.

\subsection{Ablation and Discussion}
\label{subsec:ablation}

To systematically evaluate the contribution of each architectural component to safety, interpretability, and generalization, we conducted a series of ablation studies focusing on five dimensions: the hierarchical memory structure, the hybrid decision mechanism, the unified risk encoding, the role of quantitative risk priors, and the retrieval strategy.

\textbf{Impact of Layer-2 Sub-Pattern Memory.} We first analyzed the necessity of the two-layer memory architecture. As shown in Table~\ref{tab:ablation_layer2}, removing Layer 2 resulted in a significant performance drop, particularly in complex scenarios. While the full model (L1+L2) maintained robust performance across all densities, the success rate in the most challenging setting (Lane-5, Density 3.0) fell from 70\% to 50\% when using only Layer 1, and further to 35\% with no memory. This trend highlights the strategic generalization enabled by abstract sub-patterns (e.g., ``FRONT danger $\to$ change lane''). While Layer 1 ensures precision through exact matching in familiar contexts, Layer 2 significantly expands coverage by transferring strategic knowledge to structurally similar but numerically distinct scenes, which is essential for efficient learning from sparse failures.

\begin{table}[h]
\centering
\caption{Effect of memory layers on Success Rate (\%) across different environments.}
\label{tab:ablation_layer2}
\renewcommand{\arraystretch}{1.2}
\begin{tabular}{lccc}
\hline
\textbf{Scenario} & \textbf{L1+L2 (Full)} & \textbf{L1 Only} & \textbf{No Memory} \\ \hline
Lane-4 (Density 2.0) & 100.0 & 85.0 & 70.0 \\
Lane-5 (Density 2.5) & 80.0 & 75.0 & 55.0 \\
Lane-5 (Density 3.0) & 70.0 & 50.0 & 35.0 \\ \hline
\end{tabular}
\end{table}

\textbf{Effect of Hybrid Rule--LLM Decision.} We next examined the influence of the hybrid pipeline by disabling rule-based filtering and relying solely on LLM reasoning. Without deterministic constraints, the LLM attempted unsafe lane changes in 20--30\% of high-risk cases and exhibited a 2$\times$ increase in response latency.  The hybrid mechanism leverages rules as a protective scaffold that enforces hard safety constraints (e.g., boundary limits) and filters out low-level feasibility issues, allowing the LLM to focus on higher-level, context-sensitive reasoning. This design improves both operational safety and computational efficiency.

\textbf{Utility of Structured Risk Pattern.} A comparison between RESPOND’s $5\times3$ risk pattern and raw textual scene descriptions revealed the critical role of structured encoding. When using variable-based text memory (akin to DiLu), retrieval precision suffered due to ID mismatches and numeric noise, with embedding similarity scores rarely exceeding 0.09. In contrast, the discrete risk pattern compresses irrelevant variation and aligns spatial topology with risk distribution, creating a semantically meaningful space where pattern similarity correlates directly with decision similarity. This structural consistency is what enables RESPOND to achieve deterministic retrieval and stable action reuse.

\textbf{Contribution of Quantitative Risk Values.} We also investigated the impact of explicit DRF-based risk values on zero-shot reasoning. As summarized in Table~\ref{tab:ablation_riskvalues}, removing these scalar indicators sharply reduced performance. On the real-world HighD dataset, the rate of decisions judged safer than human actions dropped from 84.9\% to 72.6\%. Similarly, the success rate in simulation fell from 70\% to 50\%. Without directional risk cues, the LLM tended to underestimate braking urgency and overestimate lateral safety. These results confirm that quantitative risk priors serve as indispensable anchors that guide the LLM toward physically grounded, risk-aware decisions in unseen conditions.

\begin{table}[h]
\centering
\caption{Contribution of directional risk values to decision quality.}
\label{tab:ablation_riskvalues}
\renewcommand{\arraystretch}{1.2}
\begin{tabular}{lcc}
\hline
\textbf{Metric} & \textbf{With Risk Value} & \textbf{Without Risk Value} \\ \hline
HighD Safer Decision Rate (\%) & 84.9& 72.6\\
Lane-4 (2.0) Success Rate (\%) & 70.0 & 50.0 \\ \hline
\end{tabular}
\end{table}

\textbf{Role of Exact Retrieval Strategy.} We evaluate the Layer~1 retrieval mechanism by replacing exact matching with approximate nearest-neighbor (ANN) search in the 15D vector space. This substitution degrades performance, as semantically dissimilar patterns can appear geometrically close under Euclidean distance. RESPOND therefore separates precision (Layer~1 exact matches) from generalization (Layer~2 abstraction), ensuring that approximation occurs only at the semantic sub-pattern level rather than the raw vector level.

In summary, the ablation results validate the cohesive design of RESPOND. Layer~2 sub-patterns enable generalization across geometrically distinct scenes, the hybrid pipeline stabilizes decision outputs, structured risk encoding ensures retrieval precision, and quantitative risk values provide essential safety priors. Together, these components support robust performance in both simulated and real-world long-tail scenarios.

\section{Conclusion}
\label{sec:conclusion}

This paper presents RESPOND, a knowledge-driven autonomous driving framework that rethinks how LLM-based agents perceive, store, and learn from risk. By encoding ego-centric risk fields into a discrete $5\times3$ matrix, RESPOND resolves the ambiguity of unstructured textual memory and establishes a foundation for precise retrieval and hybrid decision-making, effectively bridging LLM reasoning with safety-critical control requirements. Extensive experiments demonstrate that RESPOND achieves a 100\% success rate in standard highway simulations, outperforming both textual-memory baselines and reinforcement learning approaches, while exhibiting strong robustness under distribution shifts in dense traffic. Moreover, the pattern-aware reflection mechanism yields an approximately sevenfold improvement in learning efficiency, enabling the elimination of recurring failures after a single crash. Zero-shot evaluation on the HighD dataset demonstrates that RESPOND is able to generate appropriate and risk-aware responses in previously unseen, safety-critical lane-change scenarios, indicating its capability to handle real-world high-risk driving situations without prior experience. Separately, results from the personalization experiments show that RESPOND can adapt its behavior efficiently with minimal human feedback. Taken together, the above results highlight structured risk encoding and symbolic–statistical reasoning as a robust basis for generalizable and memory-efficient driving intelligence, positioning RESPOND as a promising framework for real-world safety-critical deployment.

\section*{CRediT authorship contribution statement}

Dan Chen: Conceptualization, Formal analysis, Writing – original draft, Visualization.
Heye Huang: Writing – review \& editing, Visualization, Validation.
Tiantian Chen: Data curation, Investigation.
Zheng Li: Data curation, Investigation.
Yongji Li: Formal analysis, Visualization, Investigation.
Yuhui Xu: Formal analysis, Visualization, Investigation.
Sikai Chen: Supervision, Conceptualization, Writing – review \& editing.

\section*{Declaration of competing interest}

The authors declare that they have no known competing financial interests or personal relationships that could have appeared to influence the work reported in this paper.

\section*{Data availability}

Data will be made available on request.

\bibliography{reference}

\end{document}